\documentclass[prb,twocolumn]{revtex4}
\usepackage{epsfig}
\begin{document}
\author{Dietrich Dehlinger, M. W. Mitchell}
\affiliation{~\\ Physics Department, Reed College 3203 SE Woodstock Blvd. 
\\
Portland, OR 97202}
\newcommand{\DateWritten}{\today}
\newcommand{\PutInDate}[1]{\hfill {\normalsize \rm #1}}
\email{morgan.mitchell@reed.edu}   
\title{Entangled photons, nonlocality and Bell inequalities in the 
undergraduate laboratory.}
\newcommand{\FigPos}{h} 
\newcommand{\FigWidth}{3in}
\newcommand{\SmallFigWidth}{1.75in}
\begin{abstract} 
~\\

We use polarization-entangled photon pairs to demonstrate quantum
nonlocality in an experiment suitable for advanced undergraduates. 
The photons are produced by spontaneous parametric downconversion
using a violet diode laser and two nonlinear crystals.  The
polarization state of the photons is tunable.  Using an entangled
state analogous to that described in the Einstein-Podolsky-Rosen
``paradox,'' we demonstrate strong polarization correlations of the
entanged photons.  Bell's idea of a hidden variable theory is
presented by way of an example and compared to the quantum
prediction.  A test of the Clauser, Horne, Shimony and Holt version of
the Bell inequality finds $S = 2.307 \pm 0.035$, in clear 
contradiciton of hidden variable theories.  The experiments 
described can be performed in an afternoon.
\end{abstract}
\maketitle
\newcommand{\be}{\begin{equation}}
\newcommand{\ee}{\end{equation}}
\newcommand{\bea}{\begin{eqnarray}}
\newcommand{\eea}{\end{eqnarray}}
\newcommand{\ket}[1]{\left|#1\right>}
\newcommand{\bra}[1]{\left<#1\right|}
\newcommand{\bk}{{\bf k}}
\newcommand{\etal}{{\em et al.}} 
\newcommand{\degree}{$^{\circ}$} 
\newcommand{\md}{\mbox{\degree}}

\newcommand{\HVT}{HVT}

\section{Introduction}

Entanglement of particles, an idea introduced into physics by the
famous Einstein-Podolsky-Rosen {\em Gedankenexperiment}, is one of the
most strikingly non-classical features of quantum theory.  In quantum
mechanics, particles are called entangled if their state cannot be
factored into single-particle states.  The particles are, at least in
their quantum description, inseparable.  This has surprising 
consequences. 
For example, a pair of entangled photons can show strong polarization
correlations even when each one by itself appears unpolarized.  The
standard ``Copenhagen'' interpretation of quantum measurement suggests
that these correlations arise from {\em nonlocality} of the measuring
process: a measurement on one particle instantly collapses the state
of both particles, even if they are not near each other.  Alternative
theories, which contain no such nonlocal ``action at a distance''
effects, were considered by Bell.  His ``Bell inequality'' showed 
that a very broad class of local theories disagreed with quantum 
mechanics
about the degree of polarization correlation.
Experimental tests have repeatedly found agreement with quantum 
mechanics and disagreed with this class of more intuitive, local
theories.  

Recent advances in optical technologies
have reduced the cost of producing and detecing entangled particles,
making this fascinating subject accessible to a wider audience.  Here
we describe experiments to demonstrate polarization entanglement and
test a Bell inequality.  To our knowledge, these are the first
experiments of this sort designed for undergraduates.  The new
technologies which make the experiments practical at reasonable cost,
the InGaN diode laser and the two-crystal geometry, were both
introduced in 1999.\cite{Nichia,Kwiat1999} In parallel with the
experiments we present a brief exposition of the concept of
entanglement, from its introduction by Einstein through the insights
of Bell to experimental tests.

\section{History}

Einstein remained troubled by the uncertainty principle long after
quantum mechanics had been accepted by his contemporaries.
Following a talk by Bohr in 1933,
Einstein made a comment, introducing a {\em 
Gedankenexperiment} to question the uncertainty principle.  As 
recounted by Rosenfeld, the argument was this:

``Suppose two particles are set in motion towards each other with the 
same, very large, momentum, and that they interact with each other 
for a very short time when they pass at known positions.  Consider 
now an observer who gets hold of one of the particles, far away from 
the region of interaction, and measures its momentum; then, from the 
conditions of the experiment, he will obviously be able to deduce the 
momentum of the other particle.  If, however, he chooses to measure 
the position of the first particle, he will be able to tell where the 
other particle is.  This is a perfectly correct and straightforward 
deduction from the principles of quantum mechanics; but is it not very 
paradoxical?  How can the final state of the second particle be 
influenced by a measurement performed on the first, after all 
physical interaction has ceased between them?''\cite{Rosenfeld1967}

This last sentence assumes (as Bohr had insisted) that the act of
getting information about a particle disturbs it, changing its state. 
Einstein realized that this information could be obtained by a
measurement on a different particle, with the ``paradoxical''
implication that a measurement in one place influences a
particle in another.

Two years after Bohr's talk Einstein, Podolsky and Rosen
(EPR) published a mathematical version of the same 
idea.\cite{Einstein1935}  The paper
does not suggest the ``paradoxical'' action at a distance, indeed it
assumes that such a thing is impossible.  Rather, the paper was
intended ``to expose an essential imperfection of
quantum theory.  Any attribute of a physical system that can be
accurately determined without disturbing the system, thus went the
argument, is an `element of physical reality,' and a description of
the system can only be regarded as complete if it embodies all the
elements of reality which can be attached to it.  Now, the example of
the two particles shows that the position and the momentum of a given
particle can be obtained by appropriate measurements performed on
another particle without disturbing the first, and are therefore
elements of reality in the sense indicated.  Because quantum theory does
not allow both to enter into the description of the state of the
particle, such a description is incomplete.''\cite{Rosenfeld1967}

EPR conclude their paper with a challenge of
sorts: ``While we have thus shown that the wave function does not
provide a complete description of the physical reality, we left open
the question of whether or not such a description exists.  We believe,
however, that such a theory is possible.''\cite{Einstein1935}

The ``EPR paradox,'' although it did not seriously shake confidence in
quantum mechanics, did bring to light some of its most astounding
features.  Much work has been done since then to understand the
``paradox'' that EPR raised, including both clarification of the
issues involved
and experimental tests. 
\cite{Bohr1935,Clauser1974,Peres1978,Mermin1985,
Clauser1976A,Clauser1976B,Clauser1978,Aspect1986}
Soon after the EPR paper appeared it became
clear that the ``paradox'' was not limited to position/momentum
states.  The paradoxical features remain but the math is 
simpler if we work with discrete variables such as particle spin 
or photon polarization.

\section{A polarization-entangled state}

Consider a quantum mechanical system consisting of two photons called,
for historical reasons, the ``signal'' and ``idler'' photons.  The 
photons are heading in different directions, and thus can be treated as 
distinguishable particles.  We assume the photons have the 
polarization state 
\be
\label{eq:EPRstate}
\ket{\psi_{\rm EPR}} \equiv
 \frac{1}{\sqrt{2}} 
\left(\ket{V}_{\rm s}\ket{V}_{\rm i} + 
\ket{H}_{\rm s}\ket{H}_{\rm i} \right),
\ee
where $\ket{V}$ and  $\ket{H}$ indicate vertical and horizontal 
polarization, respectively, and subscripts indicate signal or idler.
This state cannot be factored into a simple product of signal
and idler states:
$\ket{\psi_{\rm EPR}} \neq
\ket{A}_{\rm s}\ket{B}_{\rm i}$ for any choice of 
$\ket{A}_{\rm s}$ and $\ket{B}_{\rm i}$.
This means the state of one particle
cannot be specified  without making reference to the other particle. 
Such particles are  said to be ``entangled'' and $\ket{\psi_{\rm EPR}}$
is an entangled  state.

\newcommand{\unket}[1]{#1}

If we measure the polarizations of signal and idler photons in the
$H,V$ basis there are two possible outcomes: both vertical or both 
horizontal.
Each occurs half of the time.  We could instead measure the 
polarizations with polarizers rotated
by an angle $\alpha$.  We use the rotated polarization basis
\bea
\ket{V_{\alpha}} &=& \cos{\alpha}\ket{V} - \sin{\alpha}\ket{H}
\nonumber \\
\ket{H_{\alpha}} &=& \sin{\alpha}\ket{V} + \cos{\alpha}\ket{H}.
\label{eq:PolTransformation}
\eea
Here $\ket{V_{\alpha}}$ describes a state with polarization rotated by
$\alpha$ from  the vertical, while $\ket{H_{\alpha}}$ is $\alpha$ from the
horizontal. In this basis the state is
\be
\ket{\psi_{\rm EPR}} = \frac{1}{\sqrt{2}} 
\left(\ket{V_{\alpha}}_{\rm s}\ket{V_{\alpha}}_{\rm i} + 
\ket{H_{\alpha}}_{\rm s}\ket{H_{\alpha}}_{\rm i} \right).
\label{eq:EPRstateAlpha}
\ee
Clearly, if we measure in this rotated basis we get the same results:  
half the time both are $\ket{V_{\alpha}}$ and half of the time both are 
$\ket{H_{\alpha}}$.
Knowing this, we can measure the signal polarization and infer with certainty
the idler polarization.  This is the situation EPR described, but 
we've used polarizations instead of position and momentum.  Note that there 
is an uncertainty relationship between polarizations in different bases.
Knowledge of a photon's polarization in the $V_{0\md},H_{0\md}$ basis implies complete 
uncertainty of its polarization in the $V_{45\md},H_{45\md}$ basis, 
for example.

\begin{figure}[]
\centerline{\epsfig{width=\FigWidth,figure=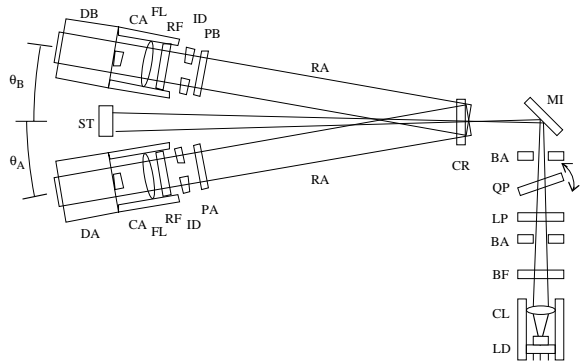}}
\label{fig:Setup} 
\renewcommand{\baselinestretch}{1}
\caption{Schematic of experimental setup, not to scale.  
Symbols:  LD Laser Diode, CL 
Collimating Lens, BF Blue Filter, BA Beam Aperture, LP Laser Polarizer, QP
Quartz Plate, MI Mirror, CR Downconversion Crystals, RA Rail, PA 
Polarizer A, PB Polarizer B, ID Iris Diaphragm, RF Red Filter,
FL Focusing Lens, CA Cage Assembly, DA Detector A, DB Detector B, ST 
Beam Stop.}
\renewcommand{\baselinestretch}{2}
\end{figure}

\section{setup}

Figure 
1
shows a schematic of our experimental setup to produce
polarization-entangled photons.  A 5 mW free-running InGaN diode laser
produces a beam of violet ($405$ nm) photons which pass through a blue
filter, a linear polarizer, and a birefringent plate 
before reaching a pair of beta barium borate (BBO) crystals.  In the 
crystals, a small
fraction of the laser photons spontaneously decay into pairs of
photons by the process of spontaneous parametric downconversion (SPD) 
.\cite{Louisell1961,Mollow1967,Giallorenzi1968,Zel'dovich1969,Hong1985,
Hariharan1996}  In a given decay the ``downconverted'' photons emerge
at the same time and on opposite sides of the laser beam.

SPD can be understood as the time-reversed process of sum-frequency
generation (SFG).  In SFG, two beams of frequency $\omega_{1}$ and
$\omega_{2}$ meet in a nonlinear crystal that lacks inversion
symmetry.  The crystal acts like a collection of ions in anharmonic
potentials.  When driven at both $\omega_{1}$ and $\omega_{2}$, the
ions oscillate with several frequency components including the sum
frequency $\omega_{1}+\omega_{2}$.  Each ion radiates at this
frequency (among others).  The coherent addition of light from each
ion in the crystal leads to constructive interference only for certain
beam directions and certain polarizations.  The condition for
constructive interference is called the ``phase matching'' requirement:
inside the crystal the wavevectors of the input beams must sum to 
that of the output beam.\cite{Boyd1992}  In SPD, the violet laser 
drives the crystal at the sum frequency and downconverted light at 
$\omega_{1}$ and $\omega_{2}$ is produced.  SPD was first used to 
test a Bell inequality in 1988. \cite{Ou1988,Shih1988}

The detectors, two single-photon counting modules (SPCMs),
are preceded by linear polarizers and red filters to block any
scattered laser light.   Even so, it is necessary to use coincidence 
detection to separate the downconverted photons from the background
of other photons reaching the detectors.  Because the photons of a 
downconverted pair are produced at the same time they cause 
coincident, i.e., nearly simultaneous, firings of the SPCMs.   
Coincidences are detected by a fast logic circuit and recorded by a
personal computer (not shown in Figure 
1
).  The detection components (SPCMs, irises, lenses and
filters) are mounted on rails which pivot about a vertical axis
passing through the crystals.  This allows the detection of SPD
photons at different angles with minimal realignment.
The setup is described in detail in the companion 
article.\cite{Dehlinger2002A}  The rails were positioned at 
$\theta_{\rm A}=\theta_{\rm B}=2.5\md$
and the focusing lenses adjusted for maximum singles rates.
With the irises fully open and polarizers both set to vertical, 
more than 300 counts per second were observed.


\begin{figure}[]
\centerline{\epsfig{width=\FigWidth,figure=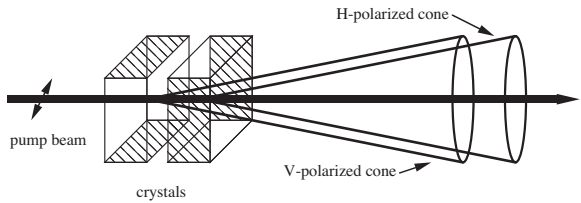}}
\label{fig:TwoCrystal} 
\renewcommand{\baselinestretch}{1}
\caption{Two-crystal downconversion source, not to scale.  The 
crystals are 0.1 mm thick and in contact face-to-face, while the 
pump beam is approximately 1 mm in diameter.  Thus the cones of 
downconverted light from the two crystals overlap almost 
completely.}
\renewcommand{\baselinestretch}{2}
\end{figure}

\section{Polarizations}

Our BBO crystals are cut for Type I phase matching, which means that 
the signal and idler photons emerge with the same polarization,
which is orthogonal to that of the pump 
photon.  Each crystal can only support downconversion 
of one pump polarization.  The other polarization passes through the 
crystal unchanged.  We use two crystals, one 
rotated 90\degree~ from the other, so that 
either pump polarization can downconvert according to the rules
\bea
\ket{V}_{p} &\rightarrow &\ket{H}_{\rm s}\ket{H}_{\rm i}
\nonumber
\\
\ket{H}_{p} &\rightarrow &\exp[i \Delta]\ket{V}_{\rm s}\ket{V}_{\rm i}
\eea
where $\Delta$ is a phase due to dispersion and birefringence in the 
crystals.  The geometry is shown schematically in Figure 2.

To create an entangled state, we first 
linearly polarize the laser beam at an angle $\theta_{l}$ from the 
vertical and
then shift the phase of one polarization component by $\phi_{l}$ 
with the birefringent quartz plate. The
laser photons (pump photons) are then in the state
\be
\ket{\psi_{\rm pump}} = \cos{\theta_{l}}\ket{V}_{p} + 
\exp[i \phi_{l}] \sin{\theta_{l}}\ket{H}_{p} 
\ee
when they reach the crystals.  The downconverted photons  
emerge in the state
\be
\ket{\psi_{\rm DC}} = \cos{\theta_{l}}\ket{H}_{\rm s}\ket{H}_{\rm i} + 
\exp[i \phi] \sin{\theta_{l}}\ket{V}_{\rm s}\ket{V}_{\rm i}
\ee
where $\phi \equiv \phi_{l}+\Delta$ is the total phase difference of 
the two polarization components.  
\cite{FNAdjustable}

\newcommand{\braket}[2]{\left<#1|#2\right>}
By placing polarizers rotated to angles $\alpha$ and $\beta$ in the 
signal and idler paths, respectively, we measure the polarization 
of the downconverted photons.  For a pair produced in the state 
$\ket{\psi_{DC}}$, the probability of coincidence detection is
\be
P_{VV}(\alpha,\beta) = 
|\bra{V_{\alpha}}_{s}\bra{V_{\beta}}_{i}\ket{\psi_{DC}}|^{2}.
\label{eq:ProbBoth}
\ee

The $VV$ subscripts on $P$ indicate the measurement outcome 
$V_{\alpha}V_{\beta}$, both photons vertical in the bases of their 
respective polarizers.  More generally, 
for any pair of polarizer angles $\alpha,\beta$, there are four 
possible outcomes, 
$V_{\alpha}V_{\beta},V_{\alpha}H_{\beta},H_{\alpha}V_{\beta}$ and 
$H_{\alpha}H_{\beta}$ indicated by $VV,VH,HV$ and $HH$, respectively.  
Using the basis of equation (\ref{eq:PolTransformation}), we 
find
\be
P_{VV}(\alpha,\beta) = \left| \sin{\alpha}\sin{\beta}\cos{\theta_{l}}
+ \exp[i\phi] \cos{\alpha}\cos{\beta}\sin{\theta_{l}}\right|^{2}
\ee
or
\bea
P_{VV}(\alpha,\beta) & = &\sin^{2}{\alpha}\sin^{2}{\beta}\cos^{2}{\theta_{l}} 
\nonumber \\
& & 
+ \cos^{2}{\alpha}\cos^{2}{\beta}\sin^{2}{\theta_{l}} 
\nonumber \\
& &
+ \frac{1}{4}\sin{2\alpha}\sin{2\beta}\sin{2\theta_{l}}\cos{\phi}
\label{eq:FirstP}
\eea
A special case occurs when 
$\ket{\psi_{DC}} = \ket{\psi_{\rm EPR}}$, i.e., when
$\theta_{l}=\pi/4$ and $\phi = 0$.  In this case 
\be
P_{VV}(\alpha,\beta) = \frac{1}{2}\cos^{2}(\beta-\alpha),
\label{eq:SpecialCase}
\ee 
which depends only on the {\em relative} angle $\beta-\alpha$.  


The last term in Equation (\ref{eq:FirstP}) is a cross term which 
accounts for the interference between the $H,H$ and $V,V$ parts of the 
state.  
The $\phi$ in this term is, through its dependence on $\Delta$, a
complicated function of pump photon wavelength, signal photon
wavelength and angle as well as crystal characteristics.  Because the 
laser has a finite linewidth and we collect
 photons over a finite solid angle and wavelength range, we
collect a range of $\phi$.  To account for this, we replace
$\cos{\phi}$ by its average $\left<\cos{\phi}\right> \equiv
\cos{\phi_{m}}$ to get
\bea
P_{VV}(\alpha,\beta) & = & \sin^{2}{\alpha}\sin^{2}{\beta}\cos^{2}{\theta_{l}} 
\nonumber \\
& & 
+ \cos^{2}{\alpha}\cos^{2}{\beta}\sin^{2}{\theta_{l}} 
\nonumber \\
& &
+ \frac{1}{4}\sin{2\alpha}\sin{2\beta}\sin{2\theta_{l}}\cos{\phi_{m}}
\eea

\newcommand{\alphabar}{{\alpha_{\perp}}}
\newcommand{\betabar}{{\beta_{\perp}}}

In our experiment we choose a fixed interval $T$ of data acquisition 
(typically in the range 0.5 seconds to 15 seconds) and record the 
number of coincidences $N(\alpha,\beta)$ during that interval.  
Assuming a constant flux of photon pairs, the number collected will
be 
\bea
N(\alpha,\beta) & = & A \left( 
\sin^{2}{\alpha}\sin^{2}{\beta}\cos^{2}{\theta_{l}} 
\right.
\nonumber \\
& & 
+ \cos^{2}{\alpha}\cos^{2}{\beta}\sin^{2}{\theta_{l}} 
\nonumber \\
& &
\left.
+ \frac{1}{4}\sin{2\alpha}\sin{2\beta}\sin{2\theta_{l}}\cos{\phi_{m}}
\right. ) + C
\label{eq:NModel}
\eea
where $A$ is the total number of entangled pairs produced
and $C$ is an offset  
to account for imperfections in the polarizers and alignment of the
crystals.  This is necessary to account for the
the fact that some coincidences are observed even when the polarizers
are set to $\alpha = 0, \beta = 90$\degree.

\section{Tuning the state}

To create the state $\ket{\psi_{\rm EPR}}$ or something close to it, 
we adjust the parameters which determine 
the laser polarization.  First we adjust $\theta_{l}$ to equalize 
the coincidence counts $N(0\md, 0\md)$ and $N(90\md , 90\md)$.  
Next we set $\phi_{l}$ by rotating the quartz plate about a 
vertical axis to maximize $N(45\md,45\md)$.  When performing 
these optimizations, we typically collect a few
hundred photons per point which requires an acquisition window of a
few seconds.


A rough idea of the purity of the entangled state can be
found by measuring $N(0\md,0\md), N(90\md,90\md), N(45\md,45\md)$ 
and $N(0\md,90\md)$.  Using the model of Equation (\ref{eq:NModel}), 
we find
\be
C = N(0\md,90\md)
\ee
\be
A = N(0\md,0\md) + N(90\md,90\md) -2C
\ee
\be
\tan^{2}\theta_{l} = \frac{N(90\md,90\md)-C}{N(0\md,0\md)-C}
\ee
\be
\cos{\phi_{m}} = \frac{1}{\sin2\theta_{l}}
\left(
4\frac{N(45\md,45\md)-C}{A} - 1
\right)
\ee


\begin{figure}[]
\centerline{\epsfig{width=\FigWidth,figure=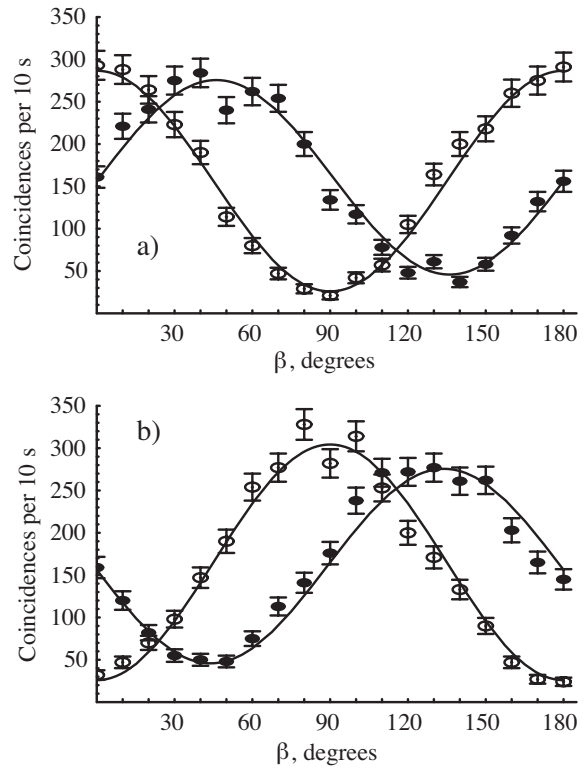}}
\label{fig:PolCorr} 
\renewcommand{\baselinestretch}{1}
\caption{Experimental polarization correlations.  a) shows $\alpha = 
0\md$ (open circles) and $\alpha = 45 \md$ (filled circles).  b) shows $\alpha = 
90\md$ (open circles) and $\alpha = 135 \md$ (filled circles).  Error 
bars indicate plus/minus one standard deviation statistical 
uncertainty.  Curves are a fit to Equation (\ref{eq:NModel}).}
\renewcommand{\baselinestretch}{2}
\end{figure}

In a typical acquisition, after optimizing $\theta_{l}$ and $\phi_{l}$
we find, with $T = 10$ seconds, $N(0,0) = 293$, $N(90,90) = 307$,
$N(0,90) = 22$, and $N(45,45) = 286$.  These give $C = 22, A = 556,
\theta_{l} = 46\md,$ and $\phi_{m} = 26\md$.  More extensive data are
shown in Figure 
3
along with a fit to Equation
(\ref{eq:NModel}).  The best fit parameters, $C = 31, A = 539,
\theta_{l} = 46\md$ and $\phi_{m} = 26\md$ are in good agreement with the
rough estimates made with just four points. 

Careful inspection of Figure 
3
shows that the theoretical curve would fit better if it were shifted
slightly to the left.  In other words, it appears as if our polarizer
angle $\beta$ is consistently off by a few degrees. 
This could be due to imperfect
positioning of the crystals, polarizers, or detector rails.  In
realignments of the lenses and rails this shift varied from 3\degree
(shown) to 8\degree, but could not be completely eliminated.  Although
we did not find it necessary to do so, a shift of this sort can be
compensated by appropriately counter-shifting the settings of $\beta$
at which measurements are taken.  This compensation has no effect on 
any of the procedures described below.

\section{Quantum measurement and entangled particles}

In his comment on Bohr's lecture, Einstein noted that quantum
mechanics allows a measurement of one particle to influence the state
of another.  To illustrate this for polarizations, we consider again
the state $\ket{\psi_{\rm EPR}}$ of Equation (\ref{eq:EPRstateAlpha}). 
If the signal photon is measured with a polarizer set to $\alpha$, the
result will be $H_{\alpha}$ or $V_{\alpha}$, each occurring half the
time.  In the usual ``Copenhagen'' interpretation the state has
collapsed, at the moment of measurement, from $\ket{\psi_{\rm EPR}}$
to either $\ket{V_{\alpha}}_{\rm s}\ket{V_{\alpha}}_{\rm i}$ or
$\ket{H_{\alpha}}_{\rm s}\ket{H_{\alpha}}_{\rm i}$.
But the mere choice
of $\alpha$ does not determine the state of the idler photon; it is the
(random) outcome of the measurement on the signal photon that decides
whether the idler ends up as $\ket{V_{\alpha}}_{i}$ or
$\ket{H_{\alpha}}_{i}$.
Despite the randomness, the choice of $\alpha$ clearly has an effect 
on the state of the idler photon: 
it gives it a definite polarization in the $\ket{V_{\alpha}}_{i},
\ket{H_{\alpha}}_{i}$ basis, which it did not have before the 
measurement.

%

The process described above is {\em nonlocal}: the state changes
instantly even though the particles could be separated by a large
distance.  We are accustomed to saying that this sort of instantaneous
action at a distance is forbidden by relativity, or that it leads to
paradoxes about sending messages to earlier times.  In this case,
though, the randomness of quantum mechanics prevents any paradoxes
from arising.  The measurement on the signal photon, whatever its
effect on the {\em state} of the idler photon, cannot be observed in
{\em measurements} on the idler photon alone.
After the signal photon is measured the idler is
equally likely to be $V_{\alpha}$ or $H_{\alpha}$.  A measurement of
its polarization, at any angle $\beta$, finds $V_{\beta}$ with 
probability
\bea
\label{eq:NoComm}
P_{V}(\beta) &=&
\frac{1}{2}|\braket{V_{\beta}}{V_{\alpha}}|^{2} +
\frac{1}{2}|\braket{V_{\beta}}{H_{\alpha}}|^{2}
\nonumber \\
&=& \frac{1}{2}[\cos^{2}(\beta-\alpha) + \sin^{2}(\beta-\alpha)]
\nonumber \\
&=& \frac{1}{2}.
\eea  
This gives no information about the choice of $\alpha$.  It is also 
the probability we would find if the signal photon had {\em not} 
been measured. 

Thus quantum mechanics (in the Copenhagen interpretation) is
consistent with relativistic causality.  It achieves that consistency
by balancing two improbable claims: the particles influence each other
nonlocally, and the randomness of nature prevents us from sending
messages that way.  A comment by Einstein succinctly
captures the oddness of this situation.  In a 1947 letter to Max Born
he objected that quantum mechanics entails ``spooky actions at a
distance.''\cite{Born1971}

\section{A Local Realistic Hidden Variable Theory}

Einstein believed that a theory could be found to replace quantum
mechanics, one which was complete and contained only local
interactions.  Here we describe such a theory, a ``local realistic
hidden variable theory'' ({\HVT}).  The name will become clear
shortly.  These were first considered by John Bell,
although our presentation most closely follows that of Aspect
.\cite{Bell1964,Bell1993,Aspect1986}  We emphasize that this theory is not a
modification of quantum mechanics (in fact it's closer to
classical mechanics).  Only the predictions of the two theories 
will be similar.

In our {\HVT}, each photon has a polarization angle $\lambda$, but this
polarization does not behave like polarization in quantum mechanics.  
When a photon meets a polarizer set to an angle $\gamma$, it will always 
register as
$V_{\gamma}$ if $\lambda$ is closer to $\gamma$ than to $\gamma+\pi/2$, 
i.e.,
\be
P_{V}^{(\HVT)}(\gamma,\lambda) = \left\{ 
\begin{array}{rccl}
    1 & & &|\gamma-\lambda| \le \pi/4 \\  
    1 & & &|\gamma-\lambda| > 3\pi/4 \\
    0 & & &{\rm otherwise.}
\end{array}
\right. 
\ee
In each pair, the signal and idler photon have the same polarization
$\lambda_{s} = \lambda_{i} = \lambda$.  As
successive pairs are produced $\lambda$ changes in an unpredictable
manner that uniformly covers the whole range of possible
polarizations.

The quantity $\lambda$ is the ``hidden variable,'' a piece of
information that is absent from quantum mechanics.  HVTs
do not have the ``spooky'' features of quantum mechanics.  The
theory is {\em local}:  measurement outcomes are determined by
features of objects present at the site of measurement.  Any
measurement on the signal (idler) photon is determined by
$\lambda_{s}$ and $\alpha$ ($\lambda_{i}$ and $\beta$).  The theory is
also {\em realistic}:  
All measurable quantities have definite values, independent of our
knowledge of them.  Furthermore, the theory specifies all of these 
values (for a given $\lambda$), so it is {\em complete} in Einstein's 
sense of the word. Finally, there is no requirement 
that $\lambda$ be random; it could be that $\lambda$ is changing 
in a deterministic way that remains to be discovered.


\begin{figure}[]
\centerline{\epsfig{width=\FigWidth,figure=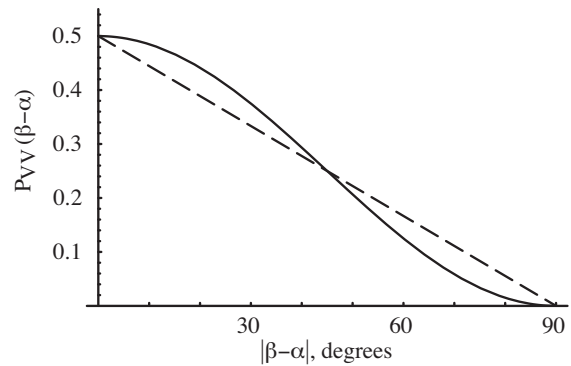}}
\label{fig:Comparison} 
\renewcommand{\baselinestretch}{1}
\caption{Predicted polarization correlations for a quantum mechanical 
entangled state (solid curve) and a hidden-variable theory (dashed 
line).}
\renewcommand{\baselinestretch}{2}
\end{figure}

To compare this theory to quantum mechanics, we need a prediction
for the coincidence probability
$P_{VV}^{({\HVT})}(\alpha,\beta)$.  A coincidence occurs when
$\lambda$ is in a range such that both $\alpha$ and $\beta$ are 
close to $\lambda$.  The probability of this is
\bea
P_{VV}^{({\HVT})}(\alpha,\beta) &=& \frac{1}{\pi}\int_{0}^{\pi}  
P_{V}^{(\HVT)}(\alpha,\lambda)P_{V}^{(\HVT)}(\beta,\lambda)d\lambda
\nonumber \\
&=&
\frac{1}{2} - \frac{|\beta-\alpha|}{\pi}.
\eea
This function and the 
corresponding quantum mechanical probability from Equation 
(\ref{eq:SpecialCase}) are plotted in Figure 
4.
The predictions are fairly similar.  Where they disagree 
quantum mechanics predicts stronger correlations (or stronger 
anti-correlations) than the \HVT.  


Our {\HVT} is very simple, and yet it agrees pretty well with quantum 
mechanics.  We might hope that some slight modification would bring
it into perfect agreement. 
In 1964 
Bell showed that this is impossible.  He derived an inequality that
all {\HVT}s obey, but which quantum mechanics violates.  We will use a
slightly different inequality, one due to Clauser, Horne, Shimony and
Holt.\cite{CHSH1969}  It is nonetheless called a Bell inequality.


\begin{figure}[]
\centerline{\epsfig{width=\SmallFigWidth,figure=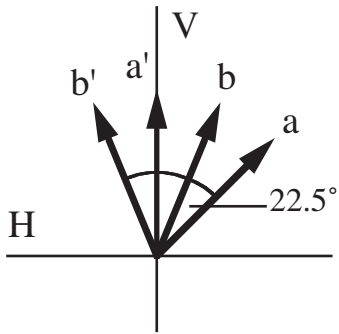}}
\label{fig:AsAndBs} 
\renewcommand{\baselinestretch}{1}
\caption{Polarizer angles for maximal $S^{(QM)}$.\hspace{\fill}} 
\renewcommand{\baselinestretch}{2}
\end{figure}

The Bell inequality constrains the degree of polarization 
correlation under 
measurements at different polarizer angles.  The proof involves 
two measures of correlation.  The first is
\bea
E(\alpha,\beta) &\equiv&P_{VV}(\alpha,\beta)+P_{HH}(\alpha,\beta)
\nonumber \\
& &
-P_{VH}(\alpha,\beta)-P_{HV}(\alpha,\beta).
\label{eq:Edef}
\eea
This incorporates all possible measurement outcomes and varies from
$+1$ when the polarizations always agree to $-1$ when they always
disagree.  The second measure is
\be
S \equiv E(a,b) - E(a,b') + E(a',b) + E(a',b'),
\label{eq:SDef}
\ee
where $a,a',b,b'$ are four different polarizer angles.  $S$ does not 
have a clear physical meaning.  Its importance comes from the
fact that Clauser, Horne, Shimony and Holt proved 
\be
| S^{} | \le 2
\label{eq:BellInequality}
\ee
for 
any {\HVT} and arbitrary $a,a',b,b'$.  
A proof is given in the Appendix.   Quantum mechanics can, for certain
settings, violate this 
inequality.  If we choose the polarizer angles, $a=-45\md, a'=0\md, 
b=22.5\md$ and $b'=22.5\md$ as shown in Figure 
5,
then, 
using eqns. (\ref{eq:SpecialCase}), (\ref{eq:Edef}), and 
(\ref{eq:SDef}),

\be
S^{(QM)} = 2 \sqrt{2}.
\label{eq:SQM}
\ee 
This result is specific to the state $\ket{\psi_{\rm EPR}}$.
Other states give lower values of $S$.
It is interesting to note that for these angles our simple {\HVT} gives
\be
S^{({\HVT})} = 2.
\ee
It mimics quantum mechanics as well as possible in light
of Equation (\ref{eq:BellInequality}).

The Bell inequality shows that no theory which is both
local and realistic (or `complete' in the EPR sense) will
ever agree with quantum mechanics.
There remains the question of whether {\em nature} agrees with quantum
mechanics or the Bell inequality.  Since we have a source that
produces photons in the state $\ket{\psi_{\rm EPR}}$, or something
close to it, we can measure $S$.  If we find $S>2$ we will have
violated the Bell inequality and thus 
disproved all {\HVT}s.  If we find $S\le 2$, no conclusion can be
drawn; both quantum mechanics and {\HVT}s are consistent with this
result.



\section{Bell inequality violation}
\label{sec:Bell}
To find the probabilities $P$ that make up $E$,
we need four values of $N$, specificially $P_{VV}(\alpha,\beta) =
N(\alpha,\beta)/N_{\rm tot}$, $P_{VH}(\alpha,\beta) =
N(\alpha,\betabar)/N_{\rm tot}$, $P_{HV}(\alpha,\beta) =
N(\alphabar,\beta)/N_{\rm tot}$ and $P_{HH}(\alpha,\beta) =
N(\alphabar,\betabar)/N_{\rm tot}$, where $N_{\rm tot}=
N(\alpha,\beta)+N(\alphabar,\beta)+
N(\alpha,\betabar)+N(\alphabar,\betabar)$ is the total number of pairs
detected and $\alphabar,\betabar$ are the polarizer
settings $\alpha+90\md, \beta+90\md$.  This requires counting
coincidences for equal intervals with the polarizers set four 
different ways.  In measuring the probabilities this way we make the
assumption that the flux of photon pairs is the same in each interval
and does not depend on the polarizer settings.  These assumptions are
reasonable, but they do create a ``loophole'' in our experimental
test.  A {\HVT}, along with the hypothesis that the polarizer settings
influence the rate of downconversion, could account for any results we
observe.  There is no evidence to support such a hypothesis.  
Nevertheless, for someone convinced of 
locality and realism, an {\em ad hoc} hypothesis of this sort may be 
more plausible than the alternative.
\cite{FNLoopholes}

%
%
%



\begin{table}[]
\begin{tabular}{rrrrrr}
    \multicolumn{1}{c}{$\alpha$} & \multicolumn{1}{c}{$\beta$} &
    \multicolumn{1}{c}{$N_{A}$}  & \multicolumn{1}{c}{$N_{B}$} & 
    \multicolumn{1}{c}{$N$} &  \multicolumn{1}{c}{$N_{\rm Ac.}$} 
    \\ \hline
   -45\degree  & -22.5\degree & 84525 & 80356 & 842 & 10.0 \\
   -45\degree  & 22.5\degree & 84607 & 82853 & 212 & 10.3 \\
 -45\degree  & 67.5\degree & 83874 & 82179 & 302 & 10.1  \\
 -45\degree  & 112.5\degree& 83769 & 77720 & 836 & 9.5  \\
 0\degree & -22.5\degree& 87015 & 80948 & 891 & 10.3  \\
 0\degree & 22.5\degree & 86674 & 83187 & 869 &    10.6 \\
 0\degree & 67.5\degree& 87086 & 81846 & 173 &    10.5 \\
 0\degree & 112.5\degree& 86745 & 77700 & 261 & 9.9 \\
 45\degree  &     -22.5\degree& 87782 & 80385 & 255 & 10.3 \\
 45\degree  & 22.5\degree& 87932 & 83265 & 830 & 10.7 \\
 45\degree  & 67.5\degree & 87794 & 81824 & 814 & 10.5 \\
 45\degree  & 112.5\degree& 88023 & 77862 & 221 & 10.1 \\
 90\degree  &     -22.5\degree& 88416 & 80941 & 170 & 10.5 \\
 90\degree  & 22.5\degree& 88285 & 82924 & 259 & 10.7 \\
 90\degree  & 67.5\degree& 88383 & 81435 & 969 & 10.6 \\
 90\degree  & 112.5\degree& 88226 & 77805 & 846 & 10.1 
\end{tabular}
\label{tab:NData}
 \caption{Singles ($N_{A},N_{B}$) and coincidence ($N$) 
 detections as a function of polarizer angles $\alpha,\beta$.  The 
 acquisition window was 
 $T = 15$ seconds, irises were fully open. Also shown are 
 ``accidental'' coincidences 
 ($N_{\rm Ac.} = \tau N_{A} N_{B} / T$) assuming a coincidence window of
$\tau = 25$ ns. }
\end{table}

A typical set of 
measurements is shown in Table I.
Also shown is the computed number of
``accidental'' coincidences, the
average number of times that photons from two different
downconversion events will arrive, purely by happenstance, within the
coincidence interval $\tau$ of each other.  This background is small,
nearly constant, and acts to decrease $|S|$.  A finding of $|S| > 2$
thus cannot be an artifact of the accidental background.

The quantity $E(\alpha,\beta)$ requires four $N$ measurements
\be
E(\alpha,\beta) = 
\frac{N(\alpha,\beta)+N(\alphabar,\betabar)
-N(\alpha,\betabar)-N(\alphabar,\beta) }
{N(\alpha,\beta)+N(\alphabar,\betabar)
+N(\alpha,\betabar)+N(\alphabar,\beta)}
\label{eq:Edef2}
\ee
and 
$S \equiv E(a,b) - E(a,b') + E(a',b) + E(a',b')$
requires sixteen.


\begin{figure}[]
\centerline{\epsfig{width=\FigWidth,figure=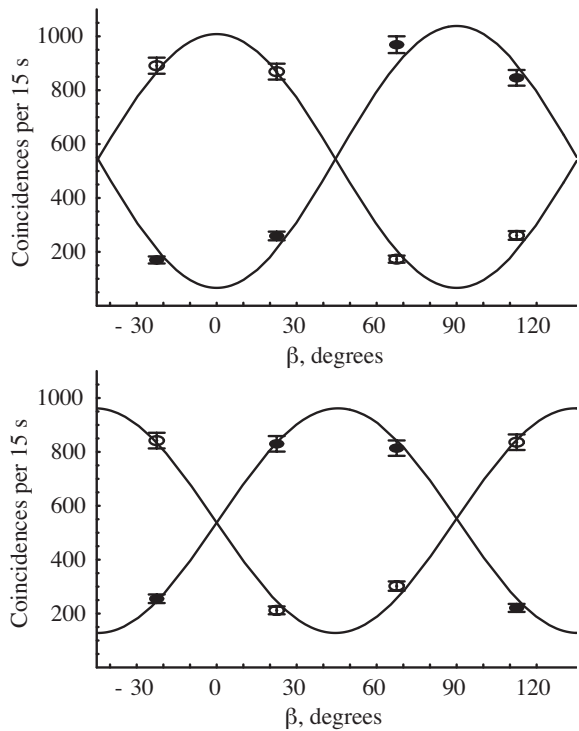}}
\label{fig:BellData} 
\renewcommand{\baselinestretch}{1}
\caption{Typical coincidence counts for the Bell inequality test.  
a) Hollow and filled circles show $\alpha = 0\md, 90\md$, respectively.
b) Hollow and filled circles show $\alpha = 45\md, 135\md$, 
respectively.  Error bars indicate plus or minus one standard deviation 
statistical uncertainty.  Curves are a fit to Equation (\ref{eq:NModel}).}
\renewcommand{\baselinestretch}{2}
\end{figure}

From these we find $S = 2.307 > 2$.  We have violated the Bell 
inequality.  To be sure of the result, we compute its 
statistical uncertainty.  The uncertainty of the $i$th measurement 
$N_{i}$  is $\sigma_{N_{i}} = \sqrt{N_{i}}$ and the 
uncertainty of the quantity $S$ is
\be
\sigma_{S} = \sqrt{\sum_{i=1}^{16} \left( 
\sigma_{N_{i}}\frac{\partial S}{\partial N_{i}}  \right)^{2}}
= \sqrt{\sum_{i=1}^{16} N_{i} \left( 
\frac{\partial S}{\partial N_{i}}  \right)^{2}}.
\ee
This innocent looking sum contains a very large number of terms
and should be evaluated by computer.  This yields 
$\sigma_{S} = 0.035$, so we have
\be
S = 2.307 \pm 0.035,
\ee
a violation of the Bell inequality by more than eight standard 
deviations.  This conclusively eliminates the {\HVT}s, and is 
consistent with quantum mechanics.  Figure 
6
shows a comparison of these data to Equation (\ref{eq:NModel}).  

\section{Interpretation}

The meaning of a Bell inequality violation is a topic for philosophy,
not experimental physics.  A good starting point for readings in the
philosophy of entanglement is a book by Michael Redhead
.\cite{Redhead1987}  Still, we will make a few comments.  These
should be understood as our (perhaps idiosyncratic) interpretation, rather than any 
consensus on the part of philosophers or physicists.

In the {\HVT}s every measurement outcome can be explained in terms of
an underlying reality in which all interactions are local.  In our
example, all possible outcomes are explained by the polarizations
$\lambda_{s}, \lambda_{i}$ of the photons, and the measurement of one
does not change the other.  In light of the Bell inequality and our
experimental findings, this sort of explanation (not just our
particular example) is impossible.  We may be able to retain one of
our assumptions, realism or locality, but not both.  Any realistic
explanation must therefore include nonlocal interactions, for example
$\lambda_{s}$ could change in response to a measurement performed on
the idler photon.  This explanation seems to be preferred by most
researchers, and an experimental Bell inequality
violation is sometimes described as a ``disproof of the principle of
locality.''  Another possibility exists: one could instead give up the
realism assumption and say that there is no underlying reality to explain
the observations, just statistical regularities relating measurement
outcomes.  If one of the goals of physics is to explain the hidden
workings of nature, accepting this position is profoundly disappointing.

It is interesting to note that a similar dilemma concerns the
interpretation of the state vector in quantum mechanics.  In the
Copenhagen interpretation, the state vector of a pair of entangled
particles changes instantaneously upon measurement.  Furthermore,
it can change in response to a measurement made on either particle,
i.e., to measurments made in different places.  If the state vector is
considered to be a real thing, then state vector collapse is an
example of instantaneous action at a distance.  But the state vector
could be viewed differently, as nothing more than a calculational
device.  After all, there is no way to measure the state vector, only
probabilities derived from it.  As shown in Equation
(\ref{eq:NoComm}), the probability for any single-particle
outcome behaves locally.  In this view, there is no action at a
distance, but there is also no answer to the question of what
``really'' is going on.

\section{Conclusion}

Using technology within reach of an undergraduate laboratory
we have created polarization-entangled photon pairs.  We have
used these to
illustrate the Einstein-Podolsky-Rosen paradox and quantum
nonlocality.  The source of entangled photons uses a violet diode 
laser and a two-crystal geometry, and can be tuned to produce
an approximation of the state $\ket{\psi_{\rm EPR}}
\equiv \left(\ket{V}_{\rm s}\ket{V}_{\rm i} + \ket{H}_{\rm
s}\ket{H}_{\rm i} \right)/\sqrt{2}$.  Polarization-sensitive
coincidence measurements clearly show the polarization correlations 
of this state, analogous to the position-momentum correlations
discussed by Einstein, Podolsky and Rosen.  Using this setup
we have shown a Bell inequality violation of more than
eight standard deviations, in clear contradiction of local realistic
hidden variable theories.

\section{acknowledgements}
We thank Paul Kwiat and David Griffiths for extensive and helpful 
comments.  This work was supported by Reed College and 
grant number DUE-0088605 from the National Science Foundation.
\appendix
\section{Proof of the CHSH Bell inequality}

For any {\HVT}, the distribution of the hidden variable $\lambda$ is 
described by a function $\rho(\lambda)$, where 
\be
\rho(\lambda) \ge 0
\ee
and
\be
\int \rho(\lambda)d\lambda  = 1.
\ee

The assumptions of locality and realism are embodied in the following:
It is assumed that for the signal photon the outcome of a measurement
is determined completely by $\lambda$ and the measurement angle
$\alpha$.  These outcomes are specified by the function
$A(\lambda,\alpha)$, which can take on the values $+1$ for detection as
$V_{\alpha}$ and $-1$ for detection as $H_{\alpha}$.  Similarly, a
function $B(\lambda,\beta)$ describes the outcomes for the idler
photon as $+1$ for $V_{\beta}$ and $-1$ for $H_{\beta}$.  A {\HVT} would
specify the functions $\rho,A$ and $B$.

The probability of a particular outcome, averaged over an ensemble
of photon pairs, is given by an integral.  In particular
\bea
P_{VV}(\alpha,\beta) &=& \int 
\frac{1+A(\lambda,\alpha)}{2}
\frac{1+B(\lambda,\beta)}{2}\rho(\lambda)d\lambda 
\nonumber \\
P_{VH}(\alpha,\beta) &=& \int 
\frac{1+A(\lambda,\alpha)}{2}
\frac{1- B(\lambda,\beta)}{2}\rho(\lambda)d\lambda 
\nonumber \\
P_{HV}(\alpha,\beta) &=& \int 
\frac{1-A(\lambda,\alpha)}{2}
\frac{1+B(\lambda,\beta)}{2}\rho(\lambda)d\lambda 
\nonumber \\
P_{HH}(\alpha,\beta) &=& \int 
\frac{1-A(\lambda,\alpha)}{2}
\frac{1-B(\lambda,\beta)}{2}\rho(\lambda)d\lambda 
\nonumber \\
\eea
are the probabilities of finding $V_{\alpha}V_{\beta}, H_{\alpha}V_{\beta}, 
H_{\alpha}V_{\beta}$ and $ H_{\alpha}H_{\beta}$, respectively.

It is easy to show that $E$, given in Equation (\ref{eq:Edef}), is
\be
E(\alpha, \beta) = \int  
A(\lambda,\alpha)B(\lambda,\beta)\rho(\lambda)d\lambda .
\ee

We define the quantity $s$, which describes the polarization 
correlation in a single pair of particles: 
\bea
s &\equiv& A(\lambda,a)B(\lambda,b)
-A(\lambda,a)B(\lambda,b')
\nonumber \\
& &
+A(\lambda,a')B(\lambda,b)
+A(\lambda,a')B(\lambda,b')
\nonumber \\
& = & A(\lambda,a)\left[B(\lambda,b) - B(\lambda,b')\right] 
\nonumber \\
& & 
+ A(\lambda,a')\left[B(\lambda,b) + B(\lambda,b')\right],
\eea
where $a,a',b,b'$ are four angles as in Equation (\ref{eq:SDef}).
Note that $s$ can only take on the values $\pm 2$.  The
average of $s$ over an ensemble of pairs is 
\bea
\left< s \right> 
&=& \int  s(\lambda,a,a',b,b')\rho(\lambda)d\lambda 
\nonumber \\
& = & 
E(a,b) - E(a,b')
+ E(a',b)  + E(a',b')
\nonumber \\
& = & S(a,a',b,b').
\eea
Because $s$ can only take on the values $\pm 2$, its average
$S$ must satisfy $-2 \le S \le +2$, which is 
the Bell inequality given in Equation (\ref{eq:BellInequality}).

%


%
%
%

\bibliographystyle{revtex}

\begin{thebibliography}{99}
\bibitem{FNAdjustable}
This state is an entangled 
state and is already quite adjustable.  Further modifications can be
made with ordinary optical components.
For example, if $\theta_{l} = \pi/4$, $\phi=\pi$ then a 
half-wave plate in the ``signal'' beam could be used to switch the 
signal polarization $\ket{H}_{s}\leftrightarrow \ket{V}_{s}$ to produce 
$
\ket{\psi_{\rm DC}'} = \left(\ket{V}_{\rm s}\ket{H}_{\rm i} - 
\ket{H}_{\rm s}\ket{V}_{\rm i} \right) /{\sqrt{2}}
$.
This is analogous to the singlet state for two spin-1/2 
particles
$\ket{\psi_{\rm singlet}}=(\ket{\uparrow}\ket{\downarrow} - 
\ket{\downarrow}\ket{\uparrow})/\sqrt{2}$.
\end{thebibliography}

\end{document}